\documentclass[aip,graphicx]{revtex4-1}
\usepackage{graphics,amsmath,amssymb,graphicx}
\usepackage{xcolor}
\draft 

\begin{document}


\title{Enhanced target normal sheath acceleration with a grooved hydrocarbon target} 



\author{Imran Khan}
\email[]{imran.skd95@gmail.com}
\affiliation{Department of physics, Indian Institute of  Technology Delhi, Hauz Khas, New Delhi, India-110016}

\author{Vikrant Saxena}
\email[]{vsaxena@physics.iitd.ac.in}

\affiliation{Department of physics, Indian Institute of  Technology Delhi, Hauz Khas, New Delhi, India-110016}

\date{\today}

\begin{abstract}
The interaction of a high-intensity ultrashort laser pulse with a  few microns-thick hydrocarbon target is known to accelerate protons/ions to multi-MeV, on the rear side of the target, via the mechanism of target normal sheath acceleration. Micro-structuring the target front is one of the promising approaches to enhance the cut-off energy as well as to reduce the divergence of accelerated protons/ions. In this paper, the interaction of a normally incident intense laser pulse with targets having single micron-sized grooves, at their front side, of semi-circular, triangular, and rectangular shapes has been studied by using two-dimensional Particle-In-Cell (PIC) simulations. It is observed that as compared to a flat target for targets with a rectangular groove at the front side the focused hot electron beam at the rear side results in an approximately four-fold increase in the cut-off energy of accelerated protons. For triangular and semi-circular groove targets, the cut-off energy remains comparatively lower (higher than the flat target though). The angular divergence of the accelerated protons/ions is also found to be relatively much lower in the case of a rectangular groove.

\end{abstract}

\maketitle 

\section{Introduction}
The plasma being an already ionized medium can support extremely high electric fields which are several orders of magnitude larger than those feasible in traditional RF accelerators. This has paved the way for the development of plasma-based compact charged particle accelerators, both for electron and ion acceleration. Laser-plasma-based acceleration of protons (and ions) to MeV energies draws motivation from their wide applications, for example, in isochoric heating of matter  \cite{patel2003isochoric}, as radiographic density diagnostic \cite{mackinnon2006proton}, for probing short-lived magnetic and electric fields in plasma (with picosecond resolution) \cite{borghesi2002electric, borghesi2003measurement}, for biomedical applications such as hadron therapy \cite{bulanov2002feasibility, ledingham2014towards}, for fast ignition of fusion targets\cite{roth2001fast, atzeni2002first}, and many more. 
To realize most of these applications a high energy conversion efficiency, from laser to protons/ions, is needed which is characterized in terms of high proton/ion cut-off energy as well as good control over beam divergence. 
In recent experiments, using high-end petawatt-class laser facilities, maximum proton energy up to 100 MeV was reported \cite{wagner2016maximum,higginson2018near}, however, with multi-terawatt lasers only a few tens of MeV proton energy is attainable which is insufficient for many potential applications.

Several laser-plasma-based ion acceleration mechanisms have been proposed in the literature, including breakout afterburner acceleration (BOA) \cite{yin2006gev, yin2011three}, relativistic-transparency acceleration \cite{badziak2006generation, gonzalez2016towards}, radiation pressure acceleration (RPA) \cite{esirkepov2004highly, robinson2008radiation}, shock acceleration\cite{silva2004proton, ji2008generating}, target normal sheath acceleration (TNSA) \cite{wilks2001energetic, snavely2000intense,mora2003plasma,passoni2008theory,passoni2010target}, etc., and a large number of investigations have been performed 
for improving the quality of accelerated ion beams. 
However, the RPA mechanism, which is operative at laser intensities $> 10^{21}$ w/cm$^{2}$, seems to be more efficient in generating better ion energies as well as lower beam divergence, it is much more demanding in terms of laser parameters. The RPA mechanism's dominance has also been observed at relatively lower intensities but with a circularly polarized laser pulse \cite{scullion2017polarization}. Typically, a very high contrast (high intensity) laser pulse is needed to avoid the deformation of ultra-thin targets/foils which are routinely used for efficient proton/ion acceleration via RPA. 


On the other hand, in  TNSA the laser pulse requirements, in terms of intensity as well as contrast, are modest. Such lasers are now routinely available and therefore TNSA has recently been widely investigated both theoretically as well as experimentally. In TNSA the laser pulse transfers its energy to the electrons near the target front surface and accelerates them in the forward direction. The hot electrons come out into the vacuum at the rear side of the target and form a sheath field (few tens of TV-m$^{-1}$) that accelerates the ions at the target rear.

However, the laser-to-ion energy conversion efficiency is less in TNSA and the accelerated ions have relatively low energy as compared to the RPA scheme.
Several strategies have been proposed to enhance the cutoff energy of protons/ions accelerated via TNSA. Their main aim has been to enhance the hot electrons' generation at the target front side. One of the possible approaches is to modify the laser pulse in order to create a pre-plasma with an optimal scale length \cite{kaluza2004influence, nuter2008influence} or to optimize the electromagnetic interference pattern \cite{ferri2019enhanced}. Another way forward is to apply an external magnetic field, which results in a converging sheath field by confining the electron trajectories (in the transverse direction) leading to higher cut-off energy of accelerated protons/ions \cite{arefiev2016enhanced, weichman2020generation}. However, the strength of the magnetic field needed is enormous and is not easily achievable. The most convenient approach to improvise the desired ion beam properties (cut-off energy, divergence, etc.) is to modify the target properties. The well-established target fabrication industry provides one the freedom to use a variety of solid TNSA targets to investigate the effect of target geometry on ion acceleration. 

There have been numerous investigations, both theoretical \cite{ferri2020enhancement,klimo2011short, feng2018effects, zou2019enhancement, andreev2011efficient, zhu2022bunched, shen2021monoenergetic, sarma2022surface} as well as experimental \cite{mackinnon2002enhancement, wagner2016maximum, cowan2004ultralow, schwoerer2006laser, floquet2013micro, purvis2013relativistic, ceccotti2013evidence, cerchez2018enhanced, qin2022high, gaillard2011increased}, addressing the effect of  target thickness\cite{mackinnon2002enhancement, wagner2016maximum}, nanostructuring of the target rear  \cite{cowan2004ultralow, schwoerer2006laser}, nanostructuring of the target front, e.g., nanoholes \cite{ferri2020enhancement},  nanocone \cite{klimo2011short, ferri2020enhancement}, nanowires \cite{feng2018effects, zou2019enhancement}, grating structure \cite{andreev2011efficient, klimo2011short}, and nanospheres \cite{klimo2011short}, etc,  on the front  side of the target, on the characteristics of the accelerated protons/ions. The structured targets produce much higher ion energies compared to the flat targets even at moderate laser power, and it strongly depends on the shape of structures as well as the angle of incidence of the laser pulse. The maximum cut-off energy of protons achieved via TNSA is reported to be $\approx$ 60 MeV in the short pulse range ($\approx$ 25 to 45 fs)\cite{ma2019laser}. Further improvement in cut-off energy and proton/ion beam quality is required to realize most of the applications.

\begin{figure}
	\includegraphics[height=4cm,width=.60\textwidth]{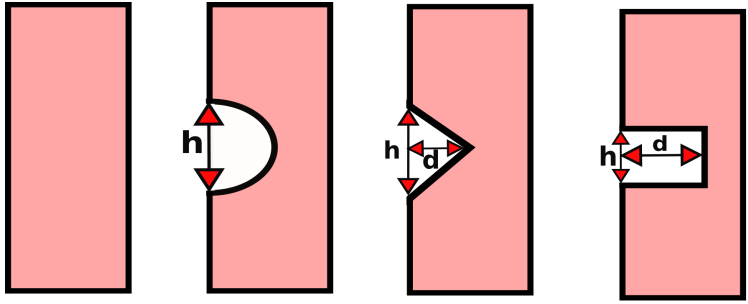}
	\caption{\label{fig:epsart} Schematic representation of the targets, without a groove (flat target), with a semi-circular groove, with a triangular groove, and with a rectangular groove, at the front side. }
    \label{fig:1}
\end{figure}

In this paper, we perform 2D particle-in-cell (PIC) simulations to investigate the effect of target front geometry on the enhancement of proton /carbon ion energies in the TNSA regime of the laser-target interaction. We consider three types of target geometries for our simulation studies, namely, targets having a rectangular groove (RG), a semi-circular groove (SCG), and a triangular groove (TG) at their front surface (as shown in Figure \ref{fig:1}), and compare them with the reference case of a flat target (FT). For each grooved target geometry, the effect of the width of the groove at a fixed depth and that of the groove depth at a fixed width has also been investigated. 

We observe an enhancement in the proton/ion cutoff energy in all three grooved target cases as compared to the flat target. The maximum cutoff energy is obtained for the target with a rectangular groove which shows approximately a four-fold increase as compared to the flat target.  This improvement in cut-off energy is a result of the effective generation of hot electrons at the front side which then travel to the rear side of the target generating an optimum sheath field that is responsible for proton/ion acceleration. In section II the simulation setup is described. Section III consists of the simulation results for single-grooved targets with different groove geometries. In section IV a detailed comparison is presented of the characteristics of ions/protons accelerated with different groove geometries on the front surface of the TNSA target, also including the flat target. This section also contains a discussion on the process of hot electron generation mechanism, and subsequent proton/ion acceleration, mainly for the most efficient rectangular groove geometry. The effect of a small-scale pre-plasma as well as of the misalignment of the laser axis is also discussed in section IV.  We also discuss the important differences our present study possesses in comparison to past studies. Finally, in section V important results of the present study are concluded.

\section{Simulation Setup}
We have performed two-dimensional PIC simulations using an open-source massively parallel PIC code, EPOCH\cite{arber2015contemporary}. 
Although it is well established in the literature that in 2D PIC simulations, the  proton/ion energies are overestimated \cite{sgattoni2012laser, d2013optimization,stark2017effects}, access to limited computational resources led us to these reduced simulations. We hope to have access to better resources in the future and will certainly verify these findings in a full-3D study.
We choose laser parameters similar to the experimental studies reported in Ref.\cite{scullion2017polarization} which were performed at GEMINI Ti:Sapphire laser at Rutherford Appleton Lab (RAL), STFC, United Kingdom. The laser pulse of wavelength $0.8\mu$m is p-polarised with intensity $5.5\times 10^{20}$  W/cm $^{2}$. The Gaussian profile is used both in space and time, with the focal spot at the waist $w_0 = 3~\mu$m and pulse duration(FWHM) =40 $fs$. 
The simulations are performed with a fully ionized quasi-neutral polyethylene ((C$_2$H$_4$)$_n$) target comprising of C$^{+6}$ ions, protons \& electrons, localized between 0 and 7 $\mu$m along the x-axis and -19.88 $\mu$m to +19.88 $\mu$m along the y-axis.
The number density of carbon ions, protons, and electrons are $4 \times10^{22}$ cm$^{-3}$, $8 \times10^{22}$ cm$^{-3}$ \& $3.2 \times10^{23}$ cm$^{-3}$, respectively. A fully ionized target has been chosen to reduce the computational cost of the simulations which is a standard practice in PIC simulations of intense laser interaction with thick overdense targets. The laser energy is more than sufficient to ionize the polyethylene target. The density of the target has been chosen corresponding to $\rho = 0.93 g/cm^3 $ (typical mass density of polyethylene) which translates to $\sim 3 \gamma n_{cr}$ (maximum value of $\gamma \sim 75$). This avoids the hole boring and collisional shock acceleration \cite{stockem2016optimizing} such that the TNSA mechanism is the dominant one. We also observe no clear signature of relativistic induced transparency (RIT)\cite{sahai2013relativistically}. In fact, due to the absence of a density gradient in our setup, the critical layer is absent. Also in the simulations performed with a pre-plasma with linear a linear density ramp, we did not observe any electron density snow-plow generated due to pondermotive force as discussed in Ref\cite{sahai2013relativistically}.

For each species, 35 macro particles per cell are used. The simulation box extends from -10 $\mu$m to 60 $\mu$m along the x-axis (i.e. direction of laser propagation) and from -20 $\mu$m to +20 $\mu$m along the y-axis. The skin depth is $l_d \approx c/w_p =9.4~nm$, where $w_p = \sqrt{n_ee^2/m_e\varepsilon _0} \:$ is plasma frequency.  The cell size is chosen as $\Delta$x = $\Delta$y = 9 $nm$  to resolve the skin depth. Thermal and open boundary conditions are used for particles in transverse and longitudinal directions respectively. Open boundaries are used for fields. All these parameters lie in the TNSA regime \cite{qiao2018revisit,higginson2018near}.

\section{Simulation Results}
The absorption mechanism $J\times B$ heating \cite{mulser2008collisionless,kruer1985j} is dominant when the laser pulse is incident normally everywhere at the front surface of the flat target whereas the vacuum heating \cite{brunel1987not,yogo2015ion} is dominant when the laser pulse is incident obliquely onto the flat target. For a structured target, part of the front surface may be subjected to effectively oblique incidence whereas the remaining part of the front surface interacts with a normally incident laser pulse. So, the vacuum heating and $J\times B$ heating mechanisms may both be simultaneously at work. Also, in the case of a structured target, the surface area for laser-plasma interaction will also increase. Hence, there is an increase in proton/ion energy. In the following, we investigate the effect of different shapes of grooves on the target front surface, on the energies and divergence of accelerated protons/ions.

\subsection{Target with a Rectangular Groove (RG)}
\begin{figure}
	\includegraphics[height=4.5cm,width=01\textwidth]{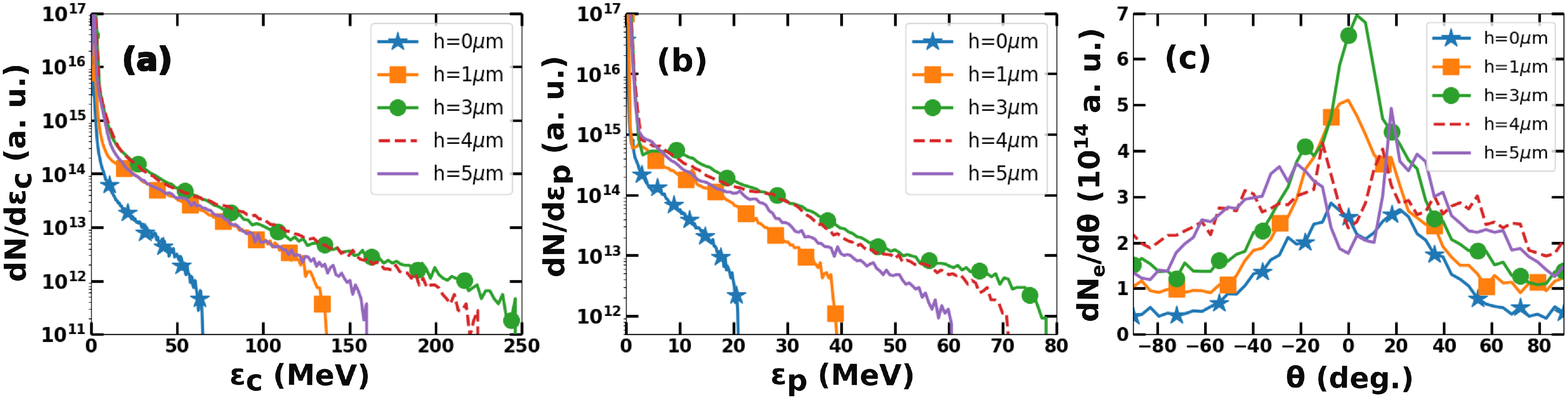}
	\caption{  (a) Carbon ions energy distribution, (b) protons energy distribution, and (c) electron angular distribution for different widths of the rectangular groove at a fixed depth ($d=5 ~\mu m$). The carbon ion and proton energy distributions are shown at $t=550~fs$ whereas the angular distribution of electrons is recorded at $t=90~fs$.}
    \label{fig:2}
\end{figure}

We first discuss the most interesting case of a target with a rectangular groove on its front surface. For this case,  we first fix the depth ($d=5 ~\mu m$) of the groove and investigate the effect of its width ($h$) by varying it between $ 1 ~\mu m$ and $5 ~\mu m$. As shown in figure \ref{fig:2} (a) \& (b), for both carbon ions and protons, on increasing the width of the groove, the cut-off energy first increases, reaches a maximum and then starts decreasing. The maximum cut-off energy for protons (and carbon ions) corresponds to the groove width $h=3 ~\mu m$ for a groove depth $ d=5 ~\mu m$.  

\begin{figure}
	\includegraphics[height=8cm,width=01\textwidth]{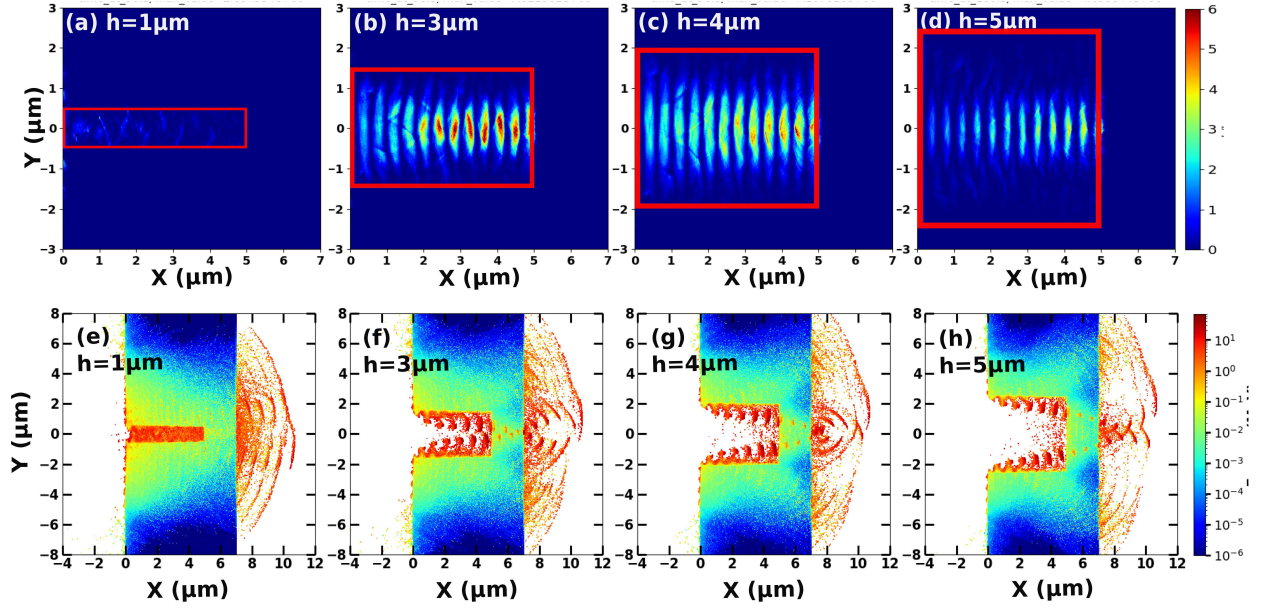}
	\caption{ 
 Laser intensity distribution (first row) at the time of peak intensity and the electron energy(MeV) distribution (second row) at time $t=70~fs$ in the xy-plane for a target with rectangular groove, having a depth $d=5~\mu m$ and a width of (a) $\&$(e) $1~\mu m$, (b) $\&$(f) $3~\mu m$, (c) $\&$(g) $4~\mu m$, and (d)$\&$(h) $5~\mu m$.}
 \label{fig:3}
\end{figure}

\begin{figure}
	\includegraphics[height=4.5cm,width=1.0\textwidth]{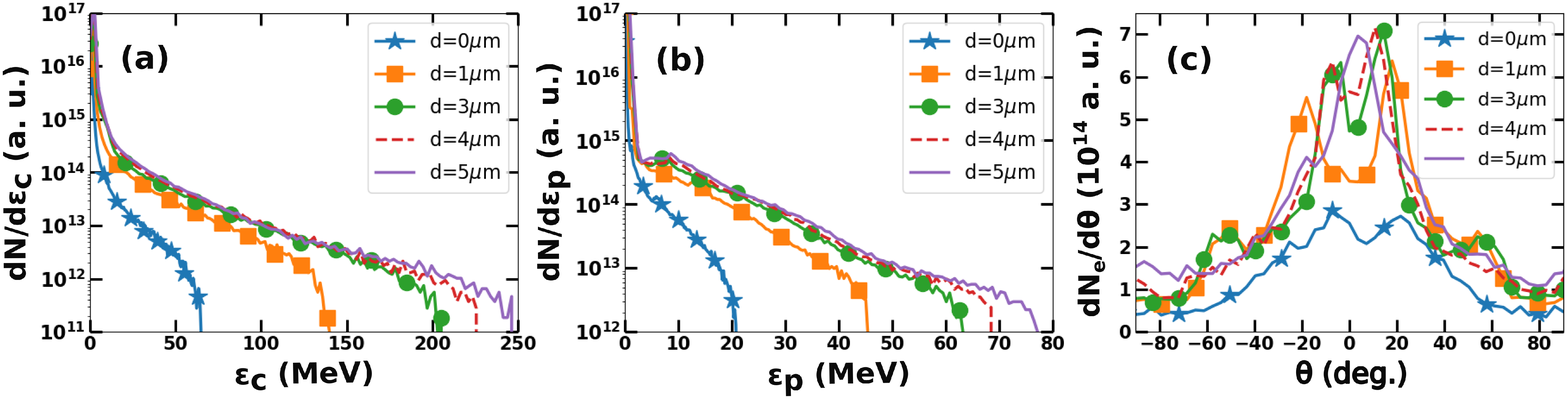}
	\caption{
 (a) Carbon ions energy distribution, (b) protons energy distribution, and (c) electron angular distribution for different depths of the rectangular groove at a fixed width ($h=5 ~\mu m$). The energy distributions for both protons and ions are  at time $t=550~fs$ whereas the angular distribution of electrons is recorded at $t=90 ~fs$.}
 \label{fig:4}
\end{figure}

The intensity of laser pulse passing through a rectangular (in 2D) aperture gets enhanced as a result of redistribution of pulse energy in the near field Fresnel region \cite{ji2016towards}. The intensification level and location highly depend on the width ($h$) of the aperture. The normalized intensification inside the groove region is defined as $\eta=I/ I_0$ where $I  \: \& \: I_0$ are the intensities in the presence and absence of the target. It is observed that as the width of the groove is increased, the intensity focusing position shifts away from the entrance of the groove, as shown in figure \ref{fig:3}(a)-(d). The magnitude of the intensification is $\eta = 2.83,\: 4.17,\: 7.02,\: 7.10,\: \& \: 4.70$ corresponding to the groove widths $h = 1~\mu m,\: 2~\mu m,\: 3~\mu m,\:4~\mu m,\: \& \: 5~\mu m,$ respectively. The cut-off energy is according to pulse intensification, except for the two cases i.e. $h=3~\mu m \: \& \: 4~\mu m$. The highest peak intensity is found to be $\eta=7.10$ which corresponds to $h=4~\mu$m in all RG geometries, although simulation results suggest that maximum cut-off energy which is $\approx 4$ times that of the flat target, corresponds to $h=3~\mu$m. 

As the laser pulse enters into the rectangular groove, the transverse component of the laser electric field pulls out a bunch of electrons from the side walls of the groove. These hot electrons couple with the laser pulse and accelerate in a forward direction by the direct laser acceleration mechanism (DLA). The target is over-dense for the selected laser parameters, the hot electrons decouple from the laser and come out at the rear side through the target and the laser pulse reflects back with reduced energy. The highly localized, high-density hot electrons form a sheath electric field at the rear side that is much higher than that caused by the bulk hot electrons described by the ponderomotive scaling \cite{wilks1992absorption} associated with TNSA.

For $h=3~\mu m$, the laser pulse is sufficiently intense in the wide range $ ~3 ~\mu m (x=2 ~\mu m $ to $5~\mu m )$. A large number of electrons are pulled out from the side walls of the groove and these hot electrons accelerate for a long time. But for $h=4\mu$m, the intensity is localized only at two points in between  $ x=4 ~\mu m $ to $5~\mu m $. Comparatively less number of electrons are pulled out from the side walls of the groove in this case and as these hot electrons are generated near the bottom of the groove they undergo acceleration for a shorter duration. It can be inferred from the electron angular distribution in figure \ref{fig:2}(c) that the highest number of focused electrons are for the case $h=3~\mu$m. The focused electron beam result in a focused sheath field and hence the protons are accelerated to the highest cut-off energy accordingly.

For fixed depth ($d=5~\mu m$) the optimized groove width is, therefore, $h=3~\mu m$. The effect of groove depth at a fixed (optimized) groove width ($h=3~\mu m$) is shown in figure \ref{fig:4}. The cut-off energy of carbon ions as well as protons increases with the increase in the groove depth. The increase in the cut-off energies is as anticipated, since the extent of the intensification region of the laser pulse increases with the groove depth, resulting in an increase in hot electron generation as well as their acceleration (by DLA) in the forward direction for a longer duration. The electrons are increasingly focused with an increase in the groove depth (seen in figure \ref{fig:4}(c)). This  results in the formation of an intense  sheath field at the rear side of the target leading to higher cut-off energy of the accelerated protons.

\subsection{Target with a Triangular Groove (TG)}
\begin{figure}
	\includegraphics[height=4.5cm,width=01\textwidth]{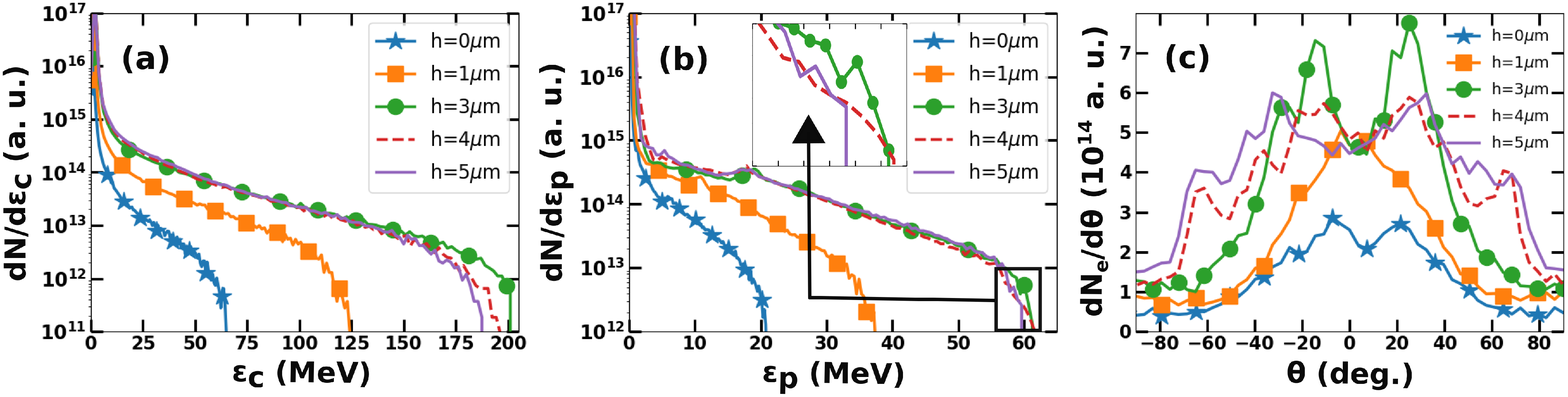}
	\caption{ (a) Carbon ions energy distribution, (b) protons energy distribution, and (c) the electron angular distribution, for different front widths at fixed depth ($d=5 ~\mu m$) of the triangular groove. The carbon ions and proton distributions are shown at time $t=550~fs$ whereas the electron angular distributions are recorded at $t=90~fs$.}
 \label{fig:5}
\end{figure}

TNSA targets with periodic and non-periodic nano-cones and nano-holes have recently been shown to enhance proton energies \cite{ferri2019enhanced}. In the present work, we consider a target with a single triangular groove of micron size and investigate the effect of its size on the cut-off energies of accelerated protons. For a fixed depth of the groove, $ d=5 ~\mu m$, the front width of the groove (or the vertex angle of the triangle) is varied. In figure \ref{fig:5}(a) $\&$(b), respectively, the energy distributions of carbon ions and protons are shown for different widths of the groove. It is noted that on increasing the front width of the groove, the cut-off energy for both carbon ions, as well as protons, first increases, reaches a maximum at $h=3~\mu m$ which is equivalent to the incident laser waist size, and then starts decreasing. 

\begin{figure}
	\includegraphics[height=4cm,width=01\textwidth]{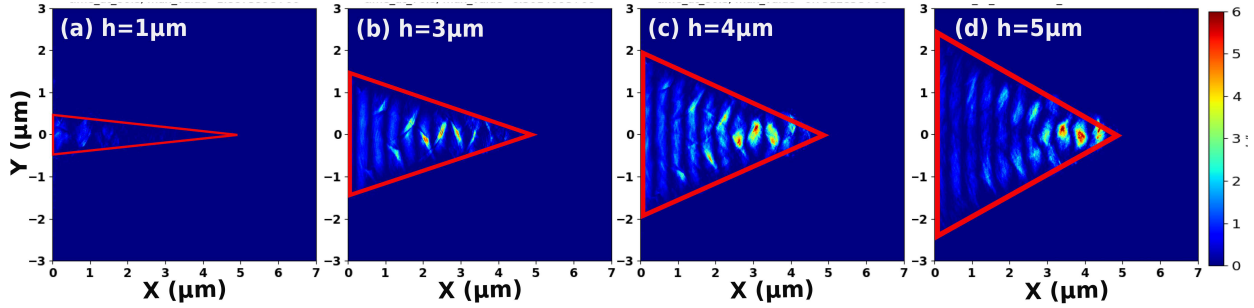}
	\caption{ Intensity distribution in the xy-plane for TG with groove widths (a) $1~\mu m$,  (b) $3~\mu m$, (c) $4~\mu m$, and (d) $5~\mu m$, at a fixed depth, $d=5~\mu m$, at the time of peak intensity.}
 \label{fig:6}
\end{figure}

To understand this behavior the pulse intensity distribution inside the groove is plotted for different groove widths in figure \ref{fig:6}(a)-(d). The maximum intensities inside the groove are $\eta = 1.88,\: 4.46,\: 6.30,\: 6.75,\: \& \: 7.20$ corresponding to the front widths $h = 1~\mu m,\: 2~\mu m,\: 3~\mu m,\:4~\mu m,\: \& \: 5~\mu m$, respectively. However, the intensification is maximum for the groove opening size $h = 4~\mu m,\: \& \: 5~\mu m$, but the cut-off energy is not maximum for these cases. As the pulse enters the groove and interacts with its slanted walls, it behaves like an obliquely incident one. It transfers the energy to the electrons much more efficiently than in a normal incidence case. The hot electrons thus produced come out in two different directions at the rear side of the target, which can be seen from the electron energy distribution in figure \ref{fig:5}(c). The hot electrons have very high divergence at the rear side for these two cases($h=4 ~\mu m$  $ \&$ $ h=5 ~\mu m$), which results in weak sheath field formation, and hence protons gain less energy. 

\begin{figure}
	\includegraphics[height=4.5cm,width=1.0\textwidth]{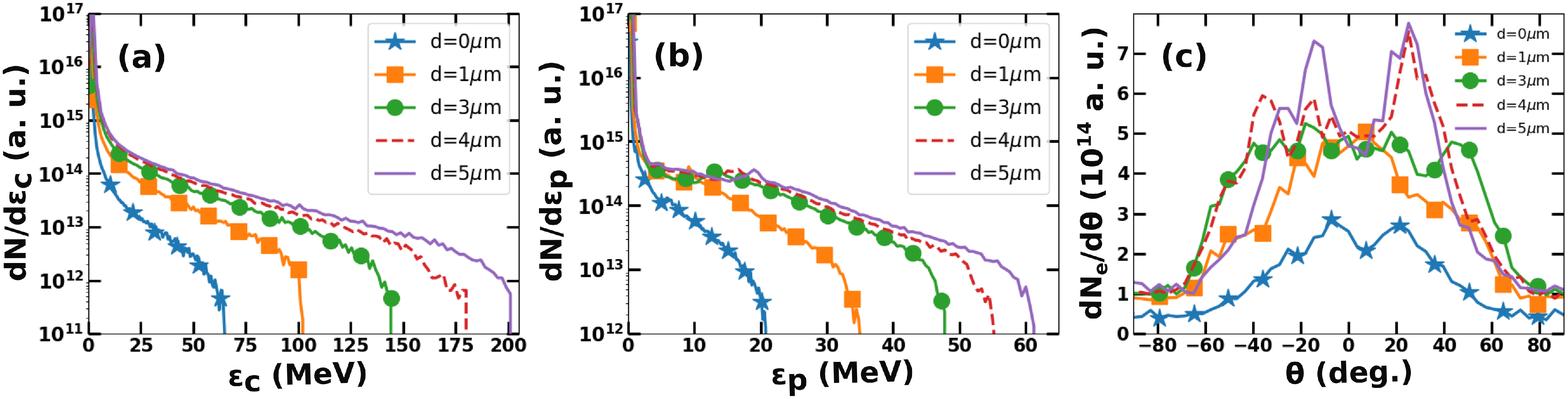}
	\caption{(a) Carbon ion energy distribution, (b) Proton energy distribution, and (c) electron angular distribution, for different groove depths at fixed groove width ($h=3 ~\mu m$) of the TG target. The carbon ion and proton energy distributions are recorded at time $t=550~fs$ whereas the electron angular distribution is at $t=90~fs$.}
 \label{fig:7}
\end{figure}

The effect of groove depth at a fixed groove width ($h=3~\mu m$) is shown in figure \ref{fig:7}. The cut-off energies of both carbon ions and protons increase with an increase in the depth of the groove. From electron angular distribution (figure \ref{fig:7}(c)), it can be noted that with increased depth the divergence of electrons at the rear side reduces which leads to the formation of a focused sheath field resulting in high cut-off energy of protons.


\subsection{Target with a Semi-Circular Groove (SCG)}
Another possible 2D geometry for a single groove on the TNSA target front is a semi-circular one. Unlike the previous two cases, in this case, the only parameter one can vary to alter the size of the groove is the radius/diameter of the groove.

\begin{figure}
	\includegraphics[height=4cm,width=01\textwidth]{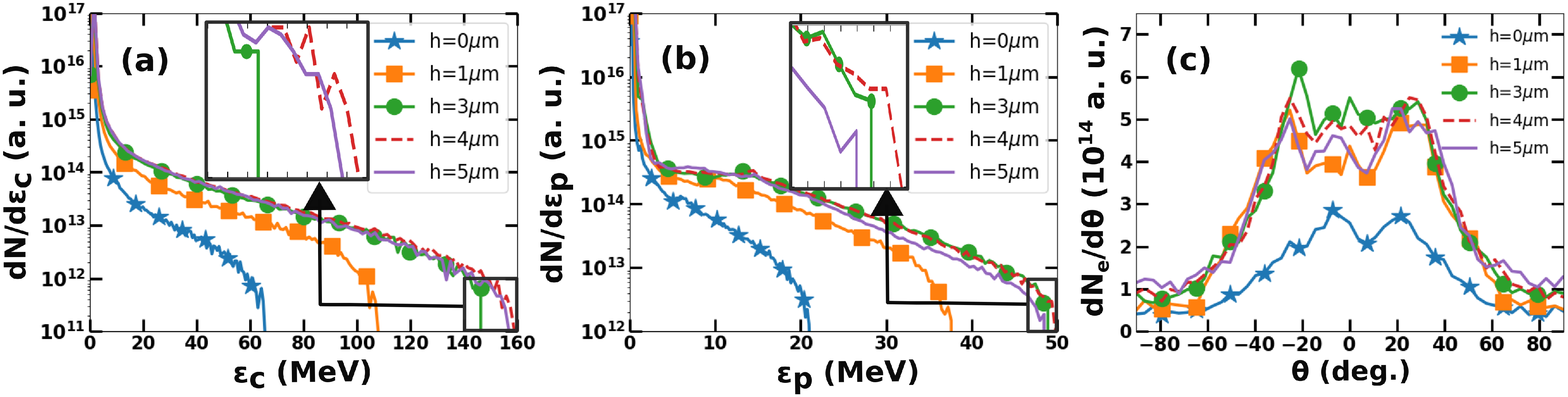}
	\caption{ (a) Carbon ion energy distribution, (b) Proton energy distribution, and (c) electron angular distribution for different groove diameters of the SCG target. The carbon ion, as well as proton energy distribution, is obtained at $t=550~fs$ whereas the angular distribution for electrons is recorded at $t=90fs$.}
 \label{fig:8}
\end{figure}

\begin{figure}
	\includegraphics[height=4cm,width=01\textwidth]{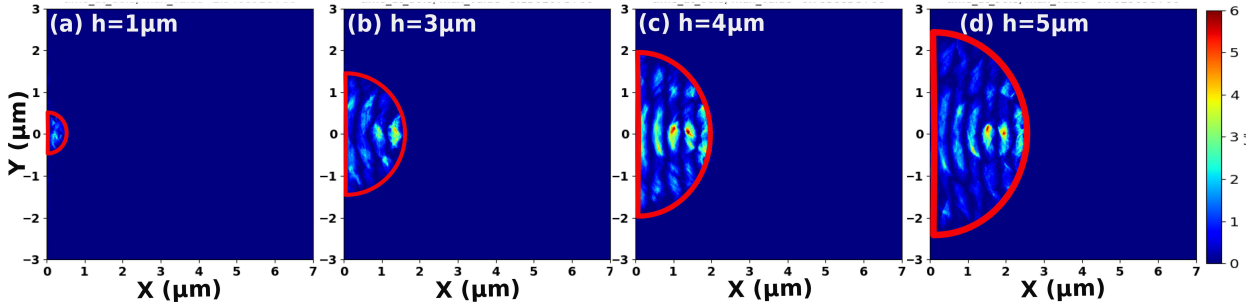}
	\caption{ Intensity distribution in the xy-plain for SCG with groove diameter (a) $1~\mu m$, (b) $3~\mu m$, (c) $4~\mu m$, and (d) $5~\mu m$, at the time of peak intensity.}
 \label{fig:9}
\end{figure}

A series of simulations are performed to investigate the effect of the diameter of the semi-circular groove on the cut-off energies of ions/protons. Figure \ref{fig:8}(a) $\&$ (b), respectively,  show the effect of the groove diameter on the energy spectra of carbon ions and protons located at the rear side of the target, at $t=550~fs$. Compared with the flat target, SCG leads to a significant gain in the cut-off energy of carbon ions and protons above 95 MeV $\&$ 28 MeV respectively. In both species, the cut-off energy first increases, reach a maximum and then decreases with an increase in groove diameter. This dependence of ion/proton cut-off energy is in accordance with pulse intensification ($\eta = 2.94,\: 4.17,\: 5.13,\: 6.78,\: \& \: 6.70$) shown in figure \ref{fig:9}. The proton cut-off energy is maximum when SCG has a groove diameter comparable to the beam waist size ($w_0\geq h=3,4~\mu m$).

The front surface of SCG is radially symmetric to the laser pulse. So one might expect that the laser pulse energy transfer to the hot electrons is maximum when the pulse coincides with the semi-circular shape. When the groove size is much less than the pulse size ($w_0\leq h=1,2~\mu m$) then most of the laser interacts with the flat part of the front surface where only the $J\times B$ mechanism work. Hence, the energy transfer from the laser pulse to the hot electrons reduces. The laser part interacts at the edges of the circular groove giving rise to two beams of hot electrons, figure \ref{fig:8}(c). The hot electrons come out at the rear side in two different directions and form a diverging sheath field. If the front opening size starts increasing beyond the pulse size, then the pulse interacts at the center of the groove and starts behaving like a flat target and absorption starts decreasing. 

\section{ Comparative Study and Discussion}
\begin{figure}
	\includegraphics[height=4.5cm,width=0.6\textwidth]{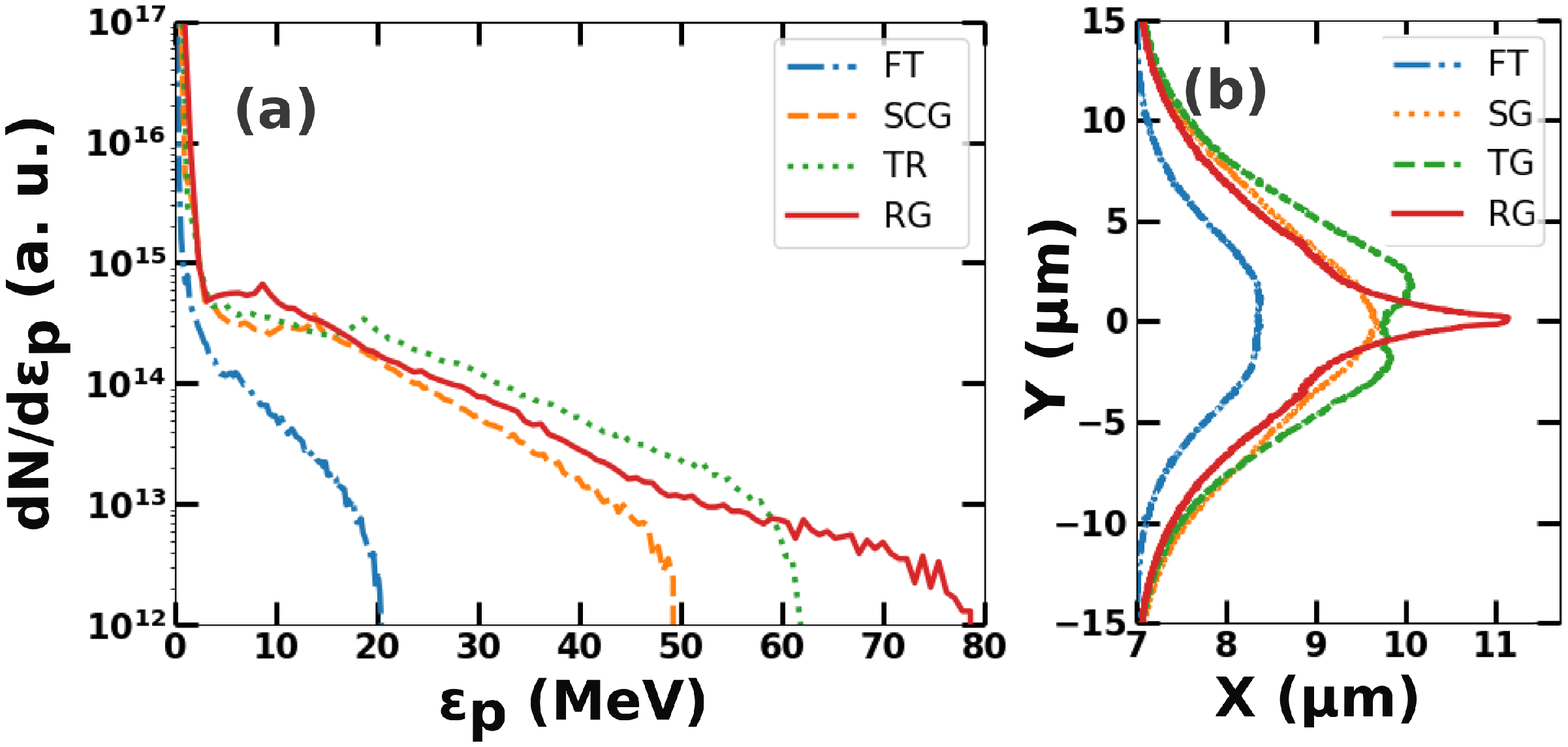}
	\caption{Comparison of proton energy distribution (a) at time $t=130~fs$, the position of the front of accelerated protons at the rear side  at time $t=130~fs$ (b). This comparison is for the case having the highest cut-off energy in each type of grooved target.}
 \label{fig:10}
\end{figure}

In the previous section, it was observed that the cut-off energy of protons is maximum when the front opening of the groove is near the beam waist size ($i.e. h=3~\mu m$) in each type of target. In this section, a comparative study is performed for the optimum cases for the three groove geometries, i.e., for groove size $h=3~\mu m\: \& \: d=5~\mu m$ in RG $\&$ TG targets, and for diameter $h=3~\mu m$ for SCG target. The proton energy spectra for these cases are compared in figure \ref{fig:10}(a). It can be noted that the RG target has the highest cut-off energy, approximately four times higher as compared with the flat target, among all the geometries. For circular and triangular groove targets, the cut-off energy is enhanced by $\approx 2.5 \: \&  \: \approx 3$ times. Also, the position of the accelerated protons at the rear side of the target is shown in figure \ref{fig:10}(b). It is apparent that the protons are much more focused in the case of a rectangular groove as compared to all other groove geometries. Similar results (not shown here) are obtained for the carbon ions as well.

These observations from the numerical simulations strongly support the rectangular shape of the groove for the efficient acceleration of a focused proton/ion beam. Although one interesting observation that can be made from the proton energy spectra is that the number of accelerated protons (in moderated energy range, 20-55 MeV), at the rear side of the target, is highest in the TG case amongst all the three groove geometries. 

\begin{figure}
	\includegraphics[height=7cm,width=01\textwidth]{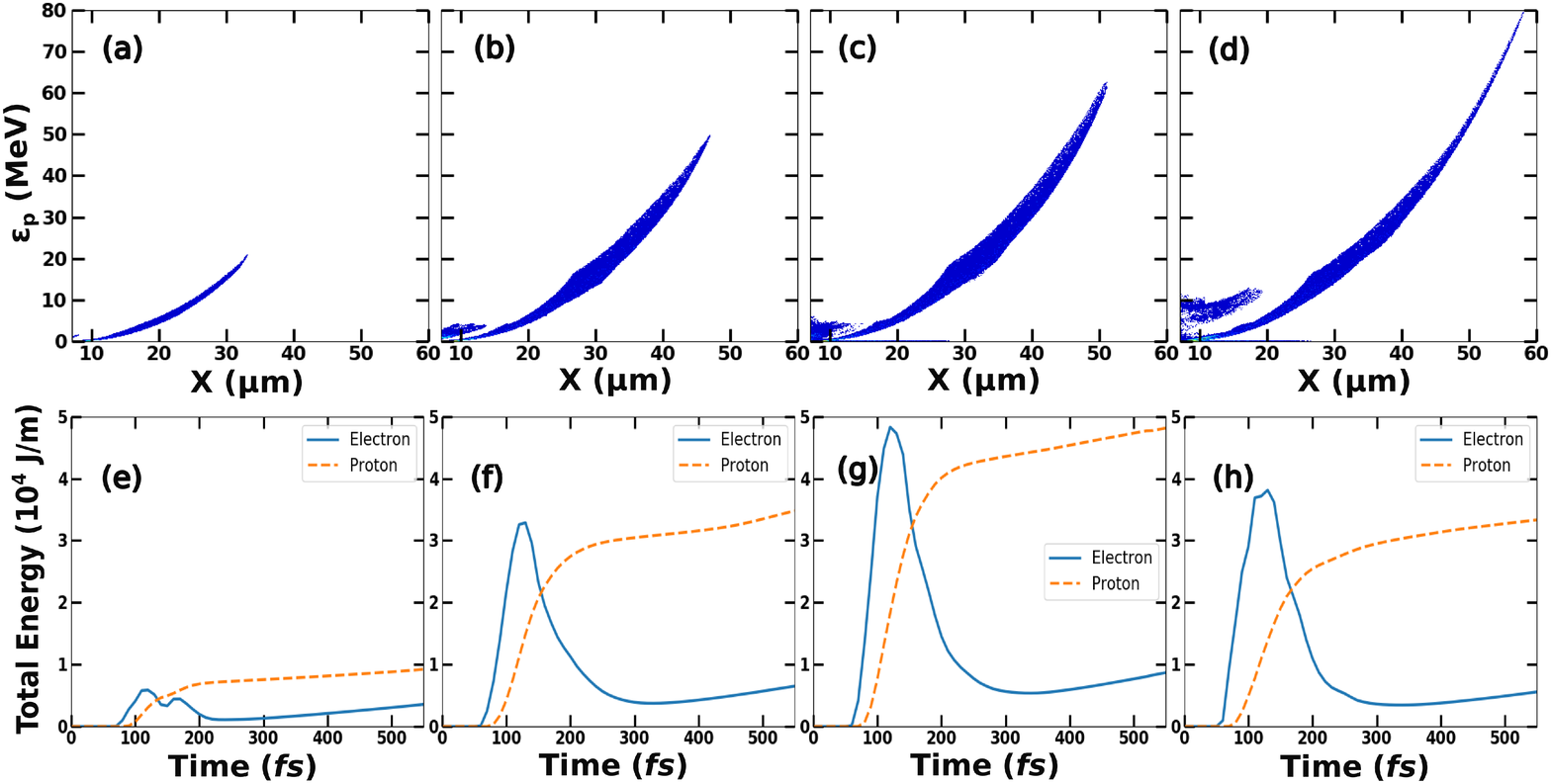}
	\caption{Proton kinetic energy distribution ( the first row) with longitudinal direction at the rear side of the target at time $t=550fs$ and time evolution of total energy (the second row) of the electron and proton at the rear side. The subfigures correspond to flat target (a) $\&$ (e), SCG (b) $\&$ (f), TG (c) $\&$ (g) and RG target (d) $\&$ (h).}
 \label{fig:11}
\end{figure}

To understand the above-described observations, the spatial distribution of the protons' kinetic energy, as well as the temporal evolution of the total kinetic energy of electrons and protons, are shown in figure \ref{fig:11}. The subplots in the first row of Figure \ref{fig:11} (subplots (a) to (d)) represent the proton kinetic energy distribution along the longitudinal direction at the rear side of the target, for each type of grooved target, at time $t=550fs$. There appear two groups of accelerated protons in the case of grooved targets, one which starts gaining energy early on and these are the ones which get highly energetic reaching up to $x=33.1 ~\mu m$, $x=47.0 ~\mu m$, $x=51.2 ~\mu m$ $\: \&$ $x=58.1 ~\mu m$ for (a) flat, (b) SCG, (c) TG, and (d) RG targets, respectively, at $t=550 ~fs$. The other population which starts gaining energy much later does not get significant energy. Evidently, the protons cover the maximum distance for the RG target case and they gain the highest acceleration and hence maximum cut-off energy ($\approx$ 80 Mev) in comparison to all other cases. The subplots in the second row of figure \ref{fig:11} represent the time evolution of the total energy of the electrons and protons at the rear side of the target. When the laser interacts with the target, it transfers part of its energy to electrons and these hot electrons transfer their energy to the protons/ions through an electrostatic sheath field. The maximum energy content of the electrons at the rear side of the target is  $0.59\times 10^4 J/m$, $3.30\times 10^4 J/m$, $4.83\times 10^4 J/m \: \& \: 3.82\times 10^4 J/m$ for (e) flat, (f) SCG, (g) TG, and (h) RG targets, respectively.
For reference, the energy content in the electromagnetic fields of the laser pulse, before interaction, is approximately $1.0 \times 10^6 J/m$.
It is very interesting to note that the total energy gained by the electrons and protons is maximum for the TG target but the highest cut-off energy is obtained for the RG target. 

\begin{figure}
	\includegraphics[height=4.5cm,width=0.6\textwidth]{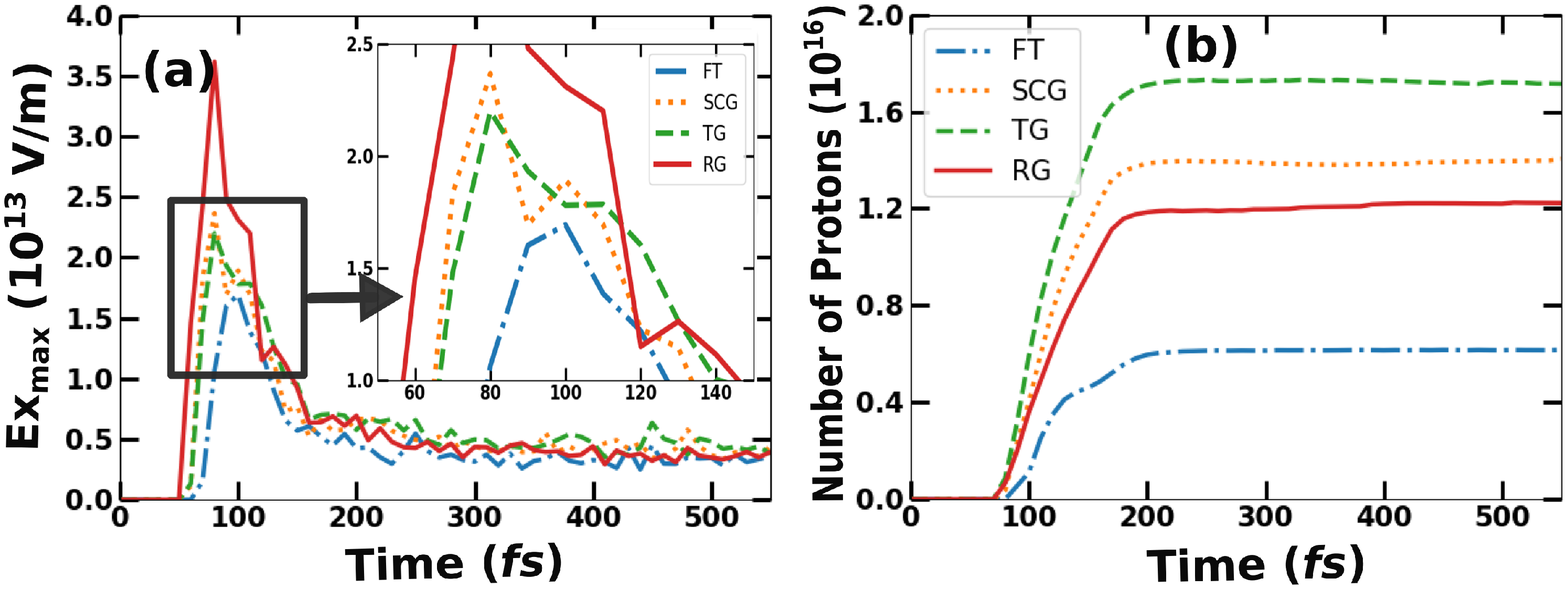}
	\caption{(a) Time evolution of lineout of the longitudinal sheath field at the rear side of the target, (b) Number of energetic protons moving at the rear side of the target having energy greater than 1.8 MeV.}
 \label{fig:12}
\end{figure}

\begin{figure}
	\includegraphics[height=4.5cm,width=0.6\textwidth]{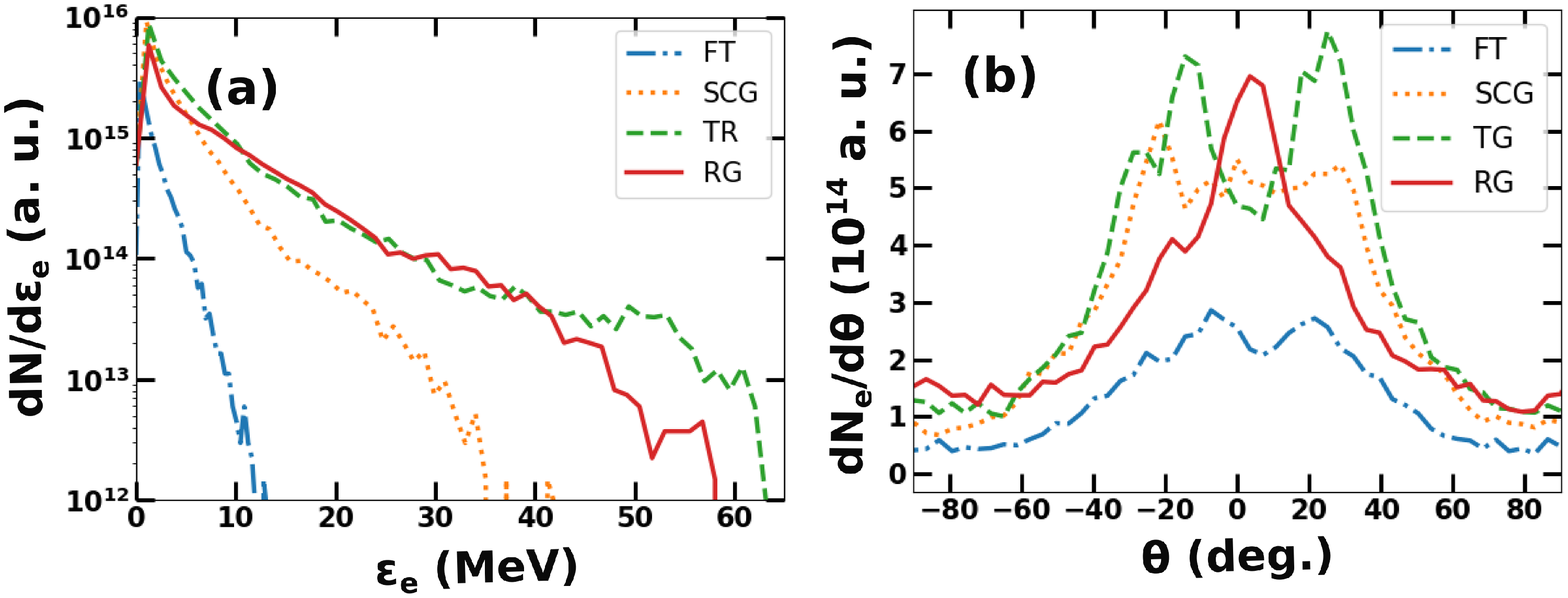}
	\caption{Electron energy spectra (a), and angular distribution (b) for different types of grooved targets at time $t=90~fs$}
 \label{fig:13}
\end{figure}

 Furthermore, figure \ref{fig:12}(a) shows the time evolution of (the lineout of) the maximum longitudinal sheath field at the rear side of the target for different target front geometries. The maximum sheath field appears around 80 fs and is 1.39, 1.30, and 2.13 times that of the flat target field for SCG, TG, and RG targets respectively. The sheath field is maximum for RG and remains for a long time compared with all other cases thus resulting in the highest cut-off energy for protons as well as carbon ions. The sheath field for SCG and TG targets remain approximately equal until $t=80 ~fs$ but at later times the sheath field remains stronger in  the TG case than in the SCG case. The electron angular distributions are shown at time $t=90~fs$, for the four cases in figure \ref{fig:13}(b). It is clearly visible from the electron's angular distributions shown in figure \ref{fig:13}(b) that for the RG target, the hot electrons are much more focused at the rear side as compared to the other two cases. Hence in the RG case, a focusing electrostatic longitudinal field is generated at the  rear side. On the other hand, the laser pulse interacting with the triangular groove has an oblique incidence onto the side walls of the groove and accelerates the hot electron in two different directions. That results in the electron distribution having two peaks which leads to a diverging electrostatic field at the rear side of the target. In the SCG target case, the hot electrons are symmetrically distributed and hence generate a highly diverging and weak field at the rear side.
 
 The populations of protons accelerated at the rear side of the target and having energy greater than 1.8MeV are plotted in figure \ref{fig:12}(b), as a function of time. The number of energetic protons is highest in the case of TG target. This is followed by the SCG target and then comes the RG target. This is consistent with the typical hot electron populations in these geometries, shown in figure \ref{fig:13}(b). 

\begin{figure}
	\includegraphics[height=4.5cm,width=.4\textwidth]{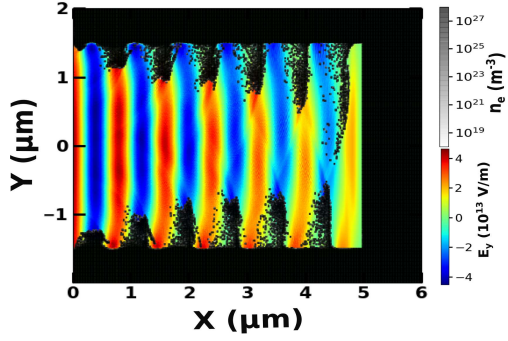}
	\caption{Spatial distribution of electron density and laser pulse electric field ($E_y$) at $t=50 ~fs$.}
 \label{fig:14}
\end{figure}

\begin{figure}
	\includegraphics[height=4.5cm,width=0.4\textwidth]{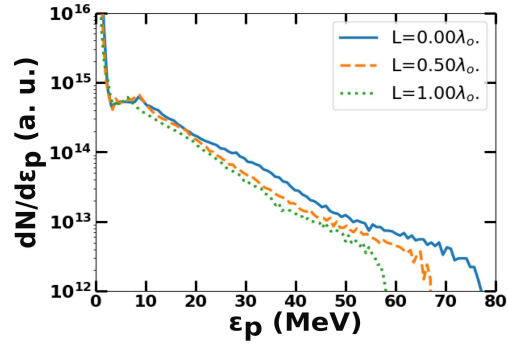}
	\caption{The change in cut-off energy of protons with an increase in the pre-plasma length at $t=550 ~fs$ }
 \label{fig:15}
\end{figure}

\begin{figure}
	\includegraphics[height=4.5cm,width=0.4\textwidth]{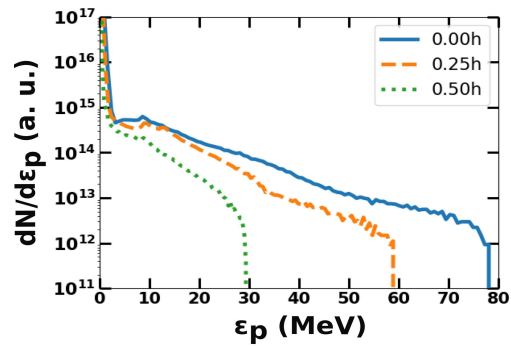}
	\caption{Effect of a mis-alignment in the laser axis on the proton energy spectra in RG case ( $t=550 ~fs$). }
 \label{fig:16}
\end{figure}

\begin{figure}
	\includegraphics[height=4.5cm,width=01\textwidth]{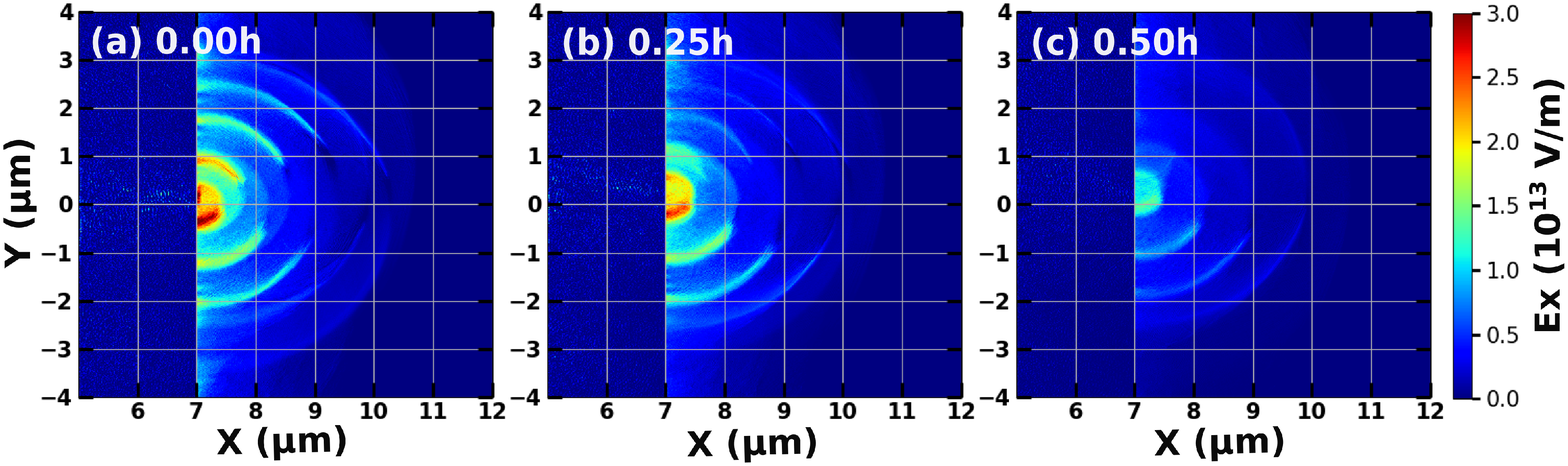}
	\caption{Effect of a mis-alignment in the laser axis on the sheath field generated at the rear side of the target.  (a) is for the without mis-alingment case and (b) $\&$ ~(c) corresponding to the mis-alingment of 0.25h $\&$ 0.5h respectively at  $t=70 ~fs$. }
 \label{fig:17}
\end{figure}

Having provided a comparison among different geometries of the groove in a 2D sense, in the following an attempt is made to understand the mechanism of laser energy absorption by the electrons inside the most effective rectangular groove. In Figure. \ref{fig:14} we show the spatial distribution of electron density as well as of the transverse laser electric field, $E_y$. It is observed in our 2D simulations that every half period of the laser field electrons from the top and the bottom walls of the groove are alternatively extracted by the transverse laser electric field, $E_y$, and are then pushed by the $\mathbf{J \times B}$ force in the forward direction. A similar energy absorption mechanism was reported by Gaillard et al. \cite{gaillard2011increased} for a peculiar $\pi$-shaped micro-cone target with a flat-top and they named it direct-laser-light-pressure acceleration (DLLPA). More recently, Zhu et al. \cite{zhu2022bunched} also reported quite similar observations for the case of a metal cone with a thin foil at its apex. Our rectangular groove case can be considered a limited case of the above two studies. In our case, the laser electric field propagates at a grazing angle along the horizontal walls of the groove with a period of $\lambda_L$ while having a phase velocity $c$. As a result, the extracted electrons are bunched with a separation of  $\lambda_L$, both along the upper as well as the lower wall of the groove while having a separation of $\lambda_L/2$ between two consecutive upper and lower electron bunches.

Clearly, having the laser electric field propagating at a grazing angle with the groove walls, the acceleration of electrons is very efficient as the transverse energy gain by the electrons from the laser electric field gets converted to their forward energy by the $\mathbf{J \times B}$ force. These electrons are then continuously forward accelerated by the laser field. Similar observations were made by Gaillard et al. \cite{gaillard2011increased} in their simulations by increasing the linear section between the curved cone and the flat top to increase the acceleration length. An enhancing  effect of the acceleration length on electron energies has also been observed in our simulations. The electron energies/temperatures are found to increase continuously as we increase the length/depth of the rectangular groove, see Figure \ref{fig:4}. On the other hand, increasing the width of the groove has an adverse effect on the hot electron energies as the DLLPA mechanism is not working at its best. It is shown in Figure \ref{fig:3}(e)-(h). The field magnitudes at the walls are relatively weaker in the case of groove width larger than the beam waist size resulting in reduced numbers as well as energies of the electrons extracted from the horizontal sides of the groove.

Similar observations of electron extraction and acceleration by the grazing laser electric field have also been reported recently by several other groups for micro-tape and wedge targets. Shen et al. \cite{shen2021monoenergetic} proposed the peeler acceleration in which a petawatt (45 fs, 50 J) laser pulse interacts with one of the shorter edges of a microtape target, exciting a surface plasma wave (SPW) along the length of the micro tape \cite{macchi2018surface}. This surface plasma wave accelerates the electrons along the length of the micro-tape leading to longitudinally bunched electrons along with a transverse focusing field. This results in a quasi-monoenergetic high energy ($>$ 100 MeV) proton beam. Inspired by their work Sarma et al. \cite{sarma2022surface} and Marini et al. \cite{marini2023electron} have also investigated, using 2D and 3D simulations, respectively, the effect of a grazing incidence of the laser pulse on a tape-like or wedge-like target and have reported an improvement in ion energies cut-off. Both these studies are similar to our case of a rectangular groove (and also the case of a triangular groove), except that in our case the SPWs are excited on both the upper and the lower walls of the groove, see Figure \ref{fig:14}.

Please note that for all the above simulations a laser pulse with an ultra-high contrast has been considered and therefore no pre-plasma is taken into account. However, this is a stringent requirement on the laser system and therefore it is of interest to understand the effect of a small-scale pre-plasma on the proton cut-off energies. To address this the effect of a small-scale pre-plasma (with linear density ramp) inside the rectangular groove, on the proton cut-off energy, has also been investigated. It is found that a pre-plasma diminishes the proton  energy cut-off which may be due to the reduced efficiency of the overall hot/fast electron generation inside the groove due to the damping of the surface plasma wave \cite{cristoforetti2020laser}. As a result, the sheath field at the rear side of the target diminishes faster than in the case of ultra-high contrast. The dependence of the proton cut-off energy on the pre-plasma scale length is shown in Figure \ref{fig:15}. We have also observed a slight forward motion of the critical layer in front of the vertical wall of the rectangular groove which may also have an effect on the electron dynamics and sheath formation \cite{sahai2014motion}.

Another important point to consider is that the above simulations have been carried out with the laser axis perfectly aligned with the target axis (along the laser propagation). However, in practice, this may be difficult to achieve and misalignment is inevitable. Keeping this in mind, a couple of simulation runs were performed with a misalignment of 0.25 h and 0.5 h in the case of a rectangular groove, where h is the width of the groove. A reduction of proton energy cut-off to a fraction of 0.24 and 0.62 is observed in the cases of the laser axis misaligned by 0.25 h and 0.5 h, respectively. 
The effect of such misalignment on the proton energy spectra is shown in Figure \ref{fig:16}.
The details on the corresponding electron energy distribution and their energy spectra are available in the supplementary material. This reduction in the proton energy cut-off is clearly the effect of reduced extraction and acceleration of electrons from both the horizontal sides of the groove which in turn results in a weaker sheath field as compared to the perfectly aligned case as shown in Figure \ref{fig:17}.

\section{Conclusions}
We have performed two-dimensional PIC simulations to study the role of the geometry of a micron-sized groove on the front surface of a hydrocarbon target, on the acceleration of carbon ions and protons, in the TNSA regime. There is a significant enhancement of proton cut-off energy in all types of groove targets in comparison with the flat target. The enhancement results from the modification in the interaction between the grooved surface and the laser pulse. The bunches of hot electrons are pulled out from the (horizontal) side walls of the groove and get accelerated to higher energies by the laser field {through a mechanism called direct laser light pressure acceleration (DLLPA). All the grooved targets have maximum energies when the front width of the groove is comparable to the laser pulse waist size enabling optimum interaction at grazing angle. The focused hot electrons at the rear side of the target result in significantly high($\approx 4 \: times$) proton cut-off energy for the target having a rectangular groove at the front side. For triangular and circular groove targets, the cut-off energy reaches only up to $\approx 3 \: \&  \: 2.5$ times due to a rather diverging sheath formation at the rear side. Moreover, both a small-scale pre-plasma, as well as a misalignment of the laser axis with respect to the target axis, seem to have a detrimental effect on the proton energy cut-off.

\section*{Supplementary Material}
See the Supplementary material for electron energy distribution of TG $\&$ RG and the effect of a misalignment in the laser axis on RG.

\begin{acknowledgments}
The authors would like to acknowledge the EPOCH consortium, for providing access to the EPOCH-4.9.0 framework \cite{arber2015contemporary}, and high-performance computing (HPC) facility at the Indian Institute of Technology Delhi for computational resources. Acknowledgments are also due to Prof. Amita Das for critically reading the manuscript. IK also acknowledges the University Grants Commission (UGC), govt. of India, for his senior research fellowship (Grant no. 1306/(CSIR-UGC NET DEC. 2018)).  
\end{acknowledgments}
\section*{Data Availability}
The data that support the findings of this study are available from the authors upon reasonable request. 
%

\end{document}